\documentclass{cimento}

\usepackage{graphicx}  

\newcommand{\grb}{GRB~051109B}

\title{Swift Observations of \grb}
\author{
E.~Troja\from{ins:DSFA}\from{ins:INAF}, G.~Cusumano\from{ins:INAF}, V.~LaParola\from{ins:INAF}, V.~Mangano\from{ins:INAF}, \atque
T.~Mineo\from{ins:INAF}
}
\instlist{
\inst{ins:DSFA} DSFA - Dipartimento di Scienze Fisiche ed Astronomiche, Palermo, Italy
\inst{ins:INAF} INAF - Istituto di Astrofisica Spaziale e Fisica Cosmica di Palermo, Italy}

\PACSes{\PACSit{98.70.Rz}{$\gamma$-ray source; $\gamma$-ray bursts}
\PACSit{01.30.Cc}{Conference proceedings}}

\begin{document}

\maketitle

\begin{abstract}
We present Swift observations of \grb, a soft long burst 
triggered by the Burst Alert Telescope (BAT).
The soft photon index of the prompt emission suggest it is a X-ray Flash (XRF) or, at least, a X-ray Rich (XRR) burst.
The X-ray lightcurve displays the canonical shape of many other GRBs, a double broken power law with a 
small flare superimposed at $\sim$T$_{\rm 0}$+1500~s, and its extrapolation to early times smoothly joins with the BAT lightcurve. 
On the basis of the derived optical to X-ray flux ratio, it cannot be classified as a dark burst.
\end{abstract}
\section{Introduction}
\begin{figure}
\centering
\label{fig:lc}
  \includegraphics[scale=0.195]{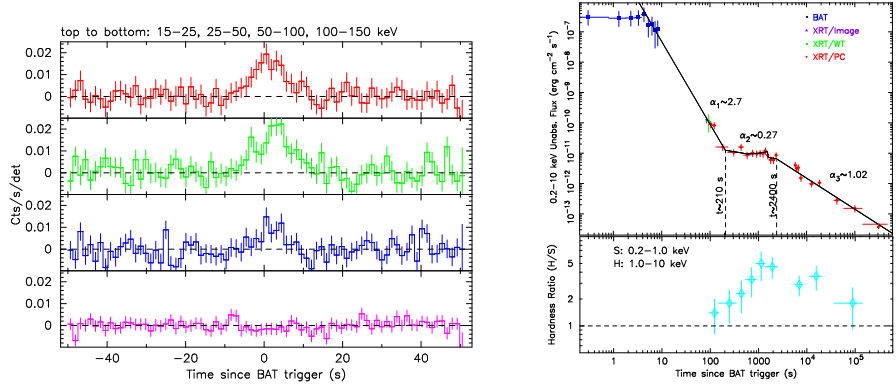}
  \caption{\small GRB~051109B lightcurves in different BAT energy bands (left)
  and  0.2--10 keV lightcurve as observed with BAT and XRT
  (right). Also shown in the lower panel the XRT/PC hardness ratio lightcurve.
}
\end{figure}

The $\gamma$-ray lightcurve of GRB~051109B shows a single soft peak,
without a strong emission above 100~keV (see Fig.~\ref{fig:lc}, left panel). 
The estimated duration in the 15--350~keV energy band is T$_{90}$=15$\pm$1~s \cite{ref:bat}.
A faint afterglow was detected by the X-Ray Telescope (XRT) $\sim$86 s after the burst and monitored for 
the following 6 days. Although the burst location was promptly observed by ground-based telescopes
(e.g. at T+52~s by the ROTSE-IIIb telescope) \cite{ref:rotse} and by the UVOT (at T+84~s) \cite{ref:first}, 
no optical counterpart was detected.

A nearby (z=0.08) barred spiral galaxy has been proposed by \cite{ref:host} as the putative host galaxy.
The field of \grb\ was then re-observed by Swift on Aug 2006 but no late-time X-ray emission was seen
in the 17~ks exposure PC image with a 3$\sigma$ upper limit of  9.8$\times$10$^{-4}$\,cts\,s$^{-1}$.

\section{Swift Data}

The time--averaged spectrum of the prompt emission is well fit by a simple power law 
with a photon index of 1.9$\pm$0.3. A cut--off powerlaw or a Band model do not improve 
significantly the fit, however fixing the first index $\alpha$ of the Band function $-1$, 
we put an upper limit to the E$_{\rm p}$ energy of 70~keV at the 90\% confidence level.
The fluence in the 15--150~keV energy band is $\sim$2.1$\times$10$^{-7}$\,erg\,cm$^{-2}$ and, using the redshift value of 0.08,
we put a lower limit of $\sim$3$\times$10$^{48}$\,erg to the isotropic energy. Extrapolating the power law model in the whole 1-10$^4$~keV
(rest frame) band we computed an upper limit of E$_{\rm iso}<$1.3$\times$10$^{49}$\,erg.
The average XRT spectrum can be modeled with an absorbed power law of \mbox{$N_H$=$(0.22\pm0.6)$}$\times 10^{22}$ cm$^{-2}$ 
and a photon index $\Gamma$=$2.1\pm0.2$. 
The fluence emitted by the afterglow in the X-ray band is $\sim$3.5$\times$10$^{-7}$\,erg\,cm$^{-2}$, comparable to that
of the prompt emission.

The combined BAT and XRT lightcurve, shown in Fig.\ref{fig:lc}, 
displays many of the features seen by Swift for other GRBs. 
The X-ray hardness ratio (Fig.\ref{fig:lc})
shows a spectral evolution between the three standard phases of the afterglow emission,
in particular a hardening related to the small flare peaking at $\sim$T+1500~s.
From spectral and timing analysis we derived a spectral index $p$ of the radiating electrons
of $\sim$2.1 for $\nu_{\rm X}$$>$$\nu_{\rm c}$.

\section{Discussion}

Due to the narrow energy band of the Swift BAT, it is hard to costrain
low peak energy values, however the soft spectral index of the prompt emission suggest GRB~051109B can 
be identified as a XRR burst.

The X-ray lightcurve has a well known shape: an initial fast decay in the first 200~s,
interpreted as off-axis prompt emission; 
a plateau, attributed to energy injection into the afterglow; 
a final decay of index $\sim$1 in agreement with standard afterglow models.
Using our best fit model and the data reported in \cite{ref:opt}, corrected for extinction, 
we found F$_X$$\sim$3.7$\times$10$^{-13}$\,erg\,cm$^{-2}$\,s$^{-1}$ and
a F$_{\rm opt}$$\sim$11 $\mu$Jy at T+11~hr. We verified that GRB~051109B
lies well above the region of dark bursts defined in \cite{ref:dark}.

\end{document}